\documentclass[aps,prd,preprint,superscriptaddress,tightenlines,nofootinbib,showpacs]{revtex4}

\usepackage{graphicx}
\usepackage{dcolumn}
\usepackage{bm}
\usepackage{amsmath}
\usepackage{amssymb}

\def \b{{\cal B}}

\begin{document}

\title{\boldmath Update of the measurement of the cross section for \\
$e^+e^-\to\psi(3770)\to\mbox{hadrons}$}

\author{D.~Besson}
\affiliation{University of Kansas, Lawrence, Kansas 66045}
\author{T.~K.~Pedlar}
\affiliation{Luther College, Decorah, Iowa 52101}
\author{D.~Cronin-Hennessy}
\author{K.~Y.~Gao}
\author{D.~T.~Gong}
\author{J.~Hietala}
\author{Y.~Kubota}
\author{T.~Klein}
\author{B.~W.~Lang}
\author{R.~Poling}
\author{A.~W.~Scott}
\author{A.~Smith}
\affiliation{University of Minnesota, Minneapolis, Minnesota 55455}
% ...
\collaboration{CLEO Collaboration} 
\noaffiliation

\date{April 7, 2010}

\begin{abstract} 

We have updated our measurement of the cross section for
$e^+e^-\to\psi(3770)\to\mbox{hadrons}$ , our publication
``Measurement of $\sigma\boldsymbol{(}
e^+e^-\to\psi(3770)\to\mbox{hadrons}\boldsymbol{)}$
at $E_{\mbox{\scriptsize{c.m.}}}=3773$~MeV'', arXiv:hep-ex/0512038,
Phys.\ Rev.\ Lett.\ {\bf 96}, 092002 (2006).
Simultaneous with this arXiv update, we have published an erratum
in Phys.\ Rev.\ Lett.\ {\bf 104}, 159901 (2010).
There, and in this update, we have
corrected a mistake in the computation of the error on the difference
of the cross sections for $e^+e^-\to\psi(3770)\to\mbox{hadrons}$ and
$e^+e^-\to\psi(3770)\to D\bar{D}$. We have also used a more recent CLEO
measurement of cross section for $e^+e^-\to\psi(3770)\to D\bar{D}$.
From this, we obtain an upper limit on the branching fraction for
$\psi(3770)\to\mbox{non-}D\bar{D}$ of $9 \%$
at $90 \%$ confidence level.

\end{abstract}

\pacs{13.25.Gv, 13.66.Bc, 14.40.-n, 99.10.Cd}
\maketitle

In our previous publication~\cite{orig}, we reported
a measurement of the cross section,
$\sigma\boldsymbol{(}e^+e^-\to\psi(3770)\to\mbox{hadrons}\boldsymbol{)}$
$\equiv$
$\sigma_{3770}= (6.38\pm0.08^{+0.41}_{-0.30})$~nb.
Then, using an earlier CLEO publication~\cite{oldddbar}, which
gave $\sigma\boldsymbol{(}e^+e^-\to\psi(3770)\to D\bar{D}\boldsymbol{)}
 \equiv \sigma_{D\bar{D}} =$
$(6.39\pm0.10^{+0.17}_{-0.08})$~nb, we obtained the cross section for
non-$D\bar{D}$ decays of $\psi(3770)$,
$\sigma_{\mbox{non}-D\bar{D}} \equiv \sigma_{3770} - \sigma_{D\bar{D}} =$
$(-0.01\pm0.08^{+0.41}_{-0.30})$~nb. 
The uncertainties on this difference made incorrect assumptions about
the correlations between the uncertainties on the two measurements.
The correct value is 
$\sigma_{\mbox{non}-D\bar{D}} = (-0.01\pm0.13^{+0.41}_{-0.33})$~nb.
In this result, the systematic uncertainty in luminosity, correlated between
the measurements of $\sigma_{3770}$ and $\sigma_{D\bar{D}}$, cancels.
The remaining systematic uncertainties, uncorrelated between the two
measurements, are combined in quadrature.

While reporting the correction to our publication, we take the
opportunity to update the measurement of $\sigma_{D\bar{D}}$ to
CLEO's latest value~\cite{newddbar},
$\sigma_{D\bar{D}} = (6.57\pm0.04\pm0.10)$~nb. With this new value,
because of different efficiencies for $D\bar{D}$ and non-$D\bar{D}$
final states, our value for $\sigma_{3770}$ changes slightly, to
$\sigma_{3770} = (6.36\pm0.08^{+0.41}_{-0.30})$~nb.
Consequently, we now have 
$\Gamma_{ee}\boldsymbol{(}\psi(3770)\boldsymbol{)} 
= (0.203\pm0.003^{+0.041}_{-0.027})$~keV.
The new value for $\sigma_{\mbox{non}-D\bar{D}}$ is 
$(-0.21\pm0.09^{+0.41}_{-0.30})$~nb.
These values supersede the previous measurements~\cite{orig}.
Dividing this difference by $\sigma_{3770}$ yields the branching
fraction $\b\boldsymbol{(}\psi(3770)\to\mbox{non-}D\bar{D}\boldsymbol{)} 
= (-3.3\pm1.4^{+6.6}_{-4.8})~\%$ which corresponds to  
$\b\boldsymbol{(}\psi(3770)\to\mbox{non-}D\bar{D}\boldsymbol{)} < 9 \%$
at $90 \%$ confidence level when considering only physical
 (positive) values.

\null

\end{document}